\documentclass[journal]{IEEEtran}

\usepackage{cite}
\usepackage{hyperref}
\usepackage{multicol}

\ifCLASSINFOpdf
   \usepackage[pdftex]{graphicx}
\else
   \usepackage[dvips]{graphicx}
\fi
\usepackage{amssymb}

\usepackage[cmex10]{amsmath}
\usepackage{algorithm}
\usepackage{multirow}

\usepackage{algorithmic}
\usepackage[tight,footnotesize]{subfigure}
\usepackage{url}
\begin{document}
\title{On-Demand Based Wireless Resources Trading for Green Communications}

\author{\IEEEauthorblockN{Wenchi Cheng$^{1}$, Xi Zhang$^{1}$, Hailin Zhang$^{2}$, and Qiang Wang$^{2}$}~\\[0.2cm]

\IEEEauthorblockA{$^{1}$Networking and Information Systems Laboratory\\
Dept. of Electrical and Computer Engineering, Texas A\&M University, College Station, TX 77843, USA\\
$^{2}$State Key Laboratory of Integrated
Services Networks, Xidian University, Xi'an, China\\
E-mail: \{\emph{wccheng@tamu.edu}, \emph{xizhang@ece.tamu.edu}, \emph{hlzhang@xidian.edu.cn}, \emph{wangqiang@gmail.com}\}}

\thanks{This work is supported by the U.S. National Science Foundation CAREER
Award under Grant ECS-0348694, the 111 Project in Xidian University of China (B08038),
National Natural Science Foundation of China (No.61072069),
the Fundamental Research Funds for the Central Universities (72101855,72105242),
the Natural Science Basic Research Plan in Shaanxi Province of China (No.2010JM8001).}
}


%Telephone: (800) 555--1212, Fax: (888) 555--1212}


\maketitle
\thispagestyle{empty}

\begin{abstract}
The purpose of Green Communications is to reduce the energy consumption of the communication system as much as possible without compromising the quality of service (QoS) for users. An effective approach for Green Wireless Communications is On-Demand strategy, which scales power consumption with the volume and location of user demand. Applying the On-Demand Communications model, we propose a novel scheme -- Wireless Resource Trading, which characterizes the trading relationship among different wireless resources for a given number of performance metrics. According to wireless resource trading relationship, different wireless resources can be consumed for the same set of performance metrics. Therefore, to minimize the energy consumption for given performance metrics, we can trade the other type of wireless resources for the energy resource under the demanded performance metrics. Based on the wireless resource trading relationship, we derive the optimal energy-bandwidth and energy-time wireless resource trading relationship for green wireless communications. We also develop an adaptive trading strategy by using different bandwidths or different delays for different transmission distances with available bandwidths and acceptable delay bounds in wireless networks. Our conducted simulations show that the energy consumption of wireless networks can be significantly reduced with our proposed wireless resources trading scheme.

\end{abstract}


\IEEEpeerreviewmaketitle

\section{Introduction}
\IEEEPARstart{R}{ecently}, the CO$_2$ emission of information and communication technologies (ICTs) has attracted a great deal of research attention~\cite{Gl,Fl,Fe}. For wireless access networks, which cost significant energy in wireless communications, a great deal of work has been done for reducing the energy consumption~\cite{Feh,Ba}. Authors of \cite{Feh} investigated the impact of deployment strategies on the power consumption of mobile radio networks. Authors of~\cite{Ba} studied the impact of reducing the cell size on the energy performance of an HSDPA RAN~\cite{HSDPA_2006}, and then proposed to save energy by putting some cells into sleep model. Authors of~\cite{Ch} proposed the energy efficient spectrum allocation for two tier cellular networks by rationally using subcarriers. Authors of~\cite{Ma} investigated the possibility of reducing the energy consumption of a cellular network by switching off some cells during the periods in which they are under-utilized when the traffic level is low. The above-mentioned works mainly explore new technologies to reduce the energy consumption of wireless access networks. However, how to minimize the minimum energy consumption of a wireless network under the demanded performance metrics has been neither well understood, nor thoroughly studied.

Clearly, to ensure scenario-specific end-user's demand, the required wireless resources need to be consumed. The required minimum wireless resources related to the On-Demand performance of users is the floor level which can guarantee the user's demand. Our studied wireless resources can be classified into the following five categories: time, bandwidth (frequency), space, energy, and code. The time resource can be seen as how long the data transmission can be delayed. The bandwidth resource can be measured as how much bandwidth can be utilized. The space resource can be considered as how many antennas can be used. The energy resource can be characterized as how much energy can be consumed. The code resource can be described as how much coding gain can be obtained by using the available codes. Most previous works focused on using the above-mentioned five categories of wireless resources for high b/s/Hz spectrum efficiency~\cite{Foschini_1998}\cite{Tarokh_1999}. These strategies show that for certain demanded performance requirements, the energy resource can be used to trade for less time and bandwidth resources consumption. Therefore, to minimize the energy consumption, we can trade time, bandwidth, space, and code resources for energy resource. In this paper, we focus on bandwidth, time, and energy domains. We derive the optimal tradeoff for minimizing energy consumption under the demanded performance metrics.

There are some related works studying the On-Demand strategy for specific wireless networks. For example, authors of~\cite{Ja} proposed a resource-on-demand policy to dynamically power on and off Wireless Local Area Network Access Points (WLAN APs), based on the volume and location of users' demand. However, they only concentrated on large-scale and high-density WLANs and switched off APs as many as possible. In contrast to~\cite{Ja}, in this paper we focus on the trading relationships among different types of wireless resources in general wireless access networks and we propose an adaptive strategy for wireless networks to consume the minimum energy while guaranteeing the demanded performance. Resource Trading can also have a wide range of networking applications, such as mobile multicast networks~\cite{XiZhang_Multicast_2002,XiZhang_Multicast_2004,XiZhang_Flow_2003}.

In this paper, we analyze the relationship between energy resource and bandwidth and time resources, respectively. We show that for certain demanded performance, less energy can be consumed at expense of consuming more other resources: bandwidth and time. Taking the operating power into account, the two type of relationships between the energy resource and the bandwidth resource and time resource, respectively, are not trivial and the claim ``the more other resource is consumed, the less energy resource is used" does not hold. Characterizing these two types of relationships, we propose an adaptive strategy to minimize the energy consumption per bit for the demanded performance at expense of consuming the bandwidth resource and time resource, respectively.

%The main purpose is to show the relation between the energy with the required performance. We do not consider any base station switching off strategy for reducing the energy consumption.

The rest of this paper is organized as follows. Section~\ref{On-Demand Communications Model} describes the model of On-Demand Communications. Section~\ref{Resource Trading in Wireless Networks} develops our Wireless Resource Trading scheme and analyzes the relationships between energy consumption per bit and bandwidth and delay time, respectively, based on which we develop an adaptive strategy on how to trade the bandwidth resource and the time resource for minimizing the energy resource, respectively. Section~\ref{Simulation Results} simulates our proposed adaptive strategy. The paper concludes with Section~\ref{Conclusions}.

%未来工作可以是：四维都考虑，时间和频率的联合考虑

\section{On-Demand Communications Model}
\label{On-Demand Communications Model}
For wireless networks, available resources can be categorized into the following five domains: time, bandwidth (frequency), space, energy and code. Maximizing QoS~\cite{Jia_TWC_2007,Cross_JSAC_XiZhang_2008,Jia_TWC_2_2007,XiZhang_Mag_2006,JiaTang_TWC_2007,Hang_TVT_2007,JiaTang_TWC_2008,Hang_CISS_2007,JiaTang_JSAC_2007} performance for wireless users implies costing more wireless resources. This is beneficial to users, but harmful to environments and operators. For environments, more energy consumption implies more CO$_{2}$ emission. For operators, more bandwidth, more time, more energy, more space, even complex code means more cost.

Therefore, a reasonable way for operators is to support the On-Demand service for users. Supporting the On-Demand service for users not only satisfies the requirement of users, but also consumes the minimum resources which may minimize the CO$_{2}$ emission and the cost of operators. In wireless networks, the model of On-Demand Communications can be expressed as follows:
\begin{equation}
\label{General-ODC}
\begin{aligned}
&\mathbf{arg min}\ Consumed\ Resources \\
&\mathrm{S.T.}\\
&User\ Obtained\ Service \geq User\ Required\ Service
\end{aligned}
\end{equation}

For green wireless networks, we limit our objective to minimize energy resource and confine our demand service to throughput and delay. Therefore, Eq.~\eqref{General-ODC} can be detailed as follows:
\begin{equation}
\begin{aligned}
&\mathbf{arg min}\ Consumed\ Energy \\
&\mathrm{S.T.}\\
&Throughput \geq User\ Required\ Throughput \\
&Delay \leq User\ Required\ Delay\ Bound \\
\end{aligned}
\end{equation}

%这里的ON-DEMAND说的少了点，后面不够了再补这里

\section{Resource Trading in Wireless Networks}
\label{Resource Trading in Wireless Networks}
It is obvious that the available resources of wireless networks are limited. The five categories of wireless resources can be consumed to satisfy the demanded users' performance.
%In the frequency, more bandwidth can这里最后扩张成期刊的时候可以加很多，
%比方从什么角度论证可以增大throughput--> shannon 公式以及一些论文，编码增益等
%但是这里为了文章的完整性，会议的文章还是要多少加一点的
Many papers have shown that using these five resources for high throughput~\cite{Foschini_1998}\cite{Tarokh_1999}, which gives us hint that certain performance can be obtained by consuming different resources. Therefore, we believe that there are potential trading relationships among the five resources. In this section, we develop the scheme of resource trading. Then, focusing on minimizing the energy resource, we derive the relationship between energy resource and bandwidth and time resource, respectively. Based on the trading relationship between energy resource and bandwidth and delay time, respectively, we develop the optimal bandwidth and delay trading strategies for green wireless networks.

\subsection{Wireless Resource Trading}
Certain performance improvement can be obtained by consuming different resources. This implies that the wireless networks can consume bandwidth, time, space, energy, and code resources for certain performance individually. Therefore, the performance improvement obtained by consuming one resource can also be obtained by consuming other types of resources. Thus there exists trading relationships among different wireless resources. Fig.~1 (a) shows the general resource trading for wireless networks. The center pentagon area represents the demanded performance of users. The other area represents the available five resources, respectively, among which the shadowing area represents the used resources for the demanded performance and the un-shadowing area represents saved resources.

For green wireless networks, the goal of resource trading is to consume the minimum energy for the demanded performance. Fig.~1(b) shows that the resource trading relationship for green wireless networks under the same demanded performance as Fig.~1(a). We can see in Fig.~1(b) that the other four resources are trading more than that in Fig.~1(a) for energy resource saving. In this paper, we mainly consider three types of resources, i.e. bandwidth, delay, energy.

\begin{center}
\begin{figure}[!htb]
\centering
\includegraphics[width=3.3 in]{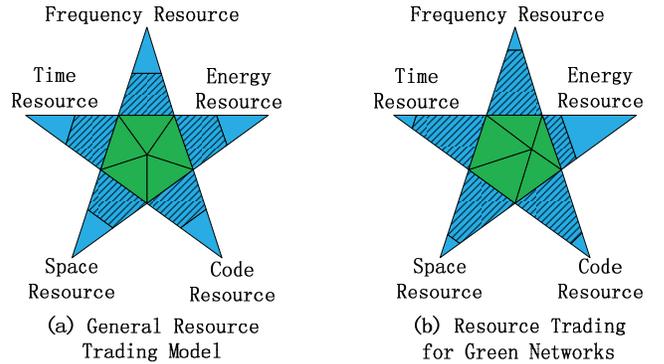}
\caption{Wireless Resources Trading Model.}
\end{figure}
\end{center}

%\begin{center}
%\begin{figure}[!htb]
%\centering
%\includegraphics[width=2.2 in]{Resource_Trading.eps}
%\caption{General Resource Trading Model}
%\end{figure}
%\end{center}
%\begin{center}
%\begin{figure}[!htb]
%\centering
%\includegraphics[width=2.2 in]{Green_Resource_Trading.eps}
%\caption{Resource Trading for Green Networks}
%\end{figure}
%\end{center}

\subsection{The Tradeoff between Energy and Bandwidth/Delay}
With capacity approaching channel codes, such as, LDPC codes, turbo codes, the data rate (channel capacity) of an AWGN channel is given by
\begin{equation}
\label{R}
\begin{aligned}
R = W\mathrm{log}_{2}\left(1+\frac{Pg}{WN_{0}}\right)
\end{aligned}
\end{equation}
where $W$ is the channel bandwidth, $P$ is the transmitted power, $g$ is the channel gain, and $N_{0}$ is the noise spectral density. The time to transmit one bit is $t$ and thus the corresponding data rate is $R = 1/t$. Thus, using Eq.\eqref{R}, we get the transmitted energy consumption per bit, denoted by $E_{tran}$, as follows:
\begin{equation}
\label{E_tran}
\begin{aligned}
E_{tran} = Pt = \frac{\left(2^{\frac{1}{Wt}}-1\right)WN_{0}t}{g}
\end{aligned}
\end{equation}
%
%Then we can obtain that the relation between transmitting power $P$ and the bandwidth $W$, delay time $t$ as
%\begin{equation}
%\begin{aligned}
%P = (2^{\cfrac{1}{Wt}}-1)WN_{0}t/g;
%\end{aligned}
%\end{equation}
From Eq.~\eqref{E_tran}, we observe that $E_{tran}$ monotonically decreases with $W$ and $t$. This is not beneficial for resource trading because not only the demanded performance cannot be satisfied, but also infinite bandwidth or time resource are needed. However, this is just for transmitting power. For green wireless networks, we must also consider operating power. Therefore, the relationship between energy resource and bandwidth resource and time resource, respectively, will be different from that of only considering transmitting power.

When considering the operating power of wireless networks, to transmit with the maximum bandwidth or the maximum delay is no longer the optimal case for minimum energy consumption. In this case, since circuit energy consumption increases with the bandwidth and the delay, we derive the overall energy consumption per bit as follows:
\begin{equation}
\label{E}
\begin{aligned}
E &= E_{tran} + E_{cir} = Pt + P_{c}t \\
  &=\frac{\left(2^{\frac{1}{Wt}}-1\right)WN_{0}t}{g} + WP_{cir}t + P_{sb}t
\end{aligned}
\end{equation}
where $E_{tran}$ and $E_{cir}$ are the energy consumption of transmitting and circuit, respectively, $P$ is the transmit power, $P_{c}$ is the circuit power, including all system power consumption except transmit power, $P_{cir}$ is the part of circuit power consumption which is related to bandwidth $W$, and $P_{sb}$ is the average static part of circuit power for every bit which is not related to $W$.

Clearly, there are two independent variables in Eq.~\eqref{E}: the bandwidth $W$ and the delay $t$, which affect the energy consumption. Ignoring the Available Resource Limitation (RAL)\footnote{RAL means the maximum available resource of the system. For example, for a wireless network, the available bandwidth is 20MHz and the acceptable delay of the data is 0.1s, then the RAL of the wireless system is 20MHz for bandwidth resource and 0.1s for time resource.} of the wireless network, the bandwidth and the time resource can trade for energy resource. Fig.~2 jointly shows the relationship by trading bandwidth resource and time resource for energy resource. Fig.~3(a) shows the partial zooming-in of Fig.~2 from $W$ = 0.3 and $t$ = 0.3 to $W$ = 0.5 and $t$ = 0.5. Fig.~3(b) shows the partial zooming-in of Fig.~2 from $W$ = 0.5 and $t$ = 0.5 to $W$ = 1, $t$ = 1. From the above two figures, we can see that the relationship between $W$ and $E$ is not monotonic. The relationship between $t$ and $E$ is not the larger $t$, the smaller $E$. TABLE~I shows the data for $t$ = 1 and $W$ = 1 in Fig.~2, respectively. In TABLE~I(a), the minimal energy consumption is 2.7725e-06. In TABLE~I(b), the minimal energy consumption is 1.4448e-06. Fig.~2, Fig.~3, and TABLE~I imply that there exists the minimum energy consumption by trading other resources. Because we want to derive the trading relationships between energy consumption and bandwidth and delay, respectively, we analyze each trading relationship with the fixed value of another resource.
\begin{center}
\begin{figure}[!htb]
\includegraphics[width=3.3 in]{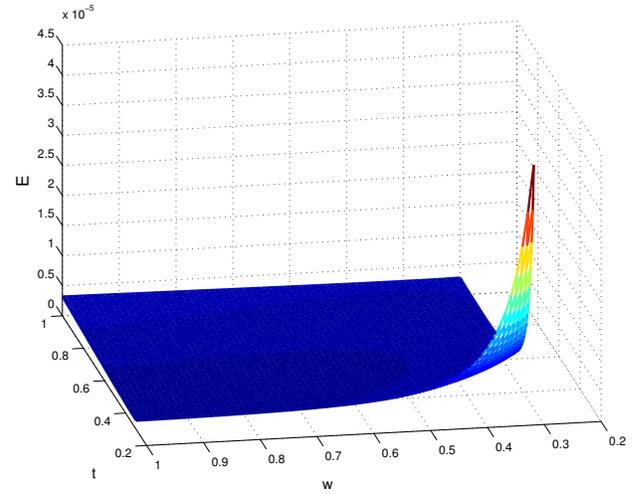}
\caption{Energy per bit versus bandwidth per bit and delay time.}
\end{figure}
\end{center}

\begin{figure}[!htb]
    \centering

    \subfigure[W = 0.3, t = 0.3 to W = 0.5, t = 0.5]{
        \includegraphics[scale=0.4]{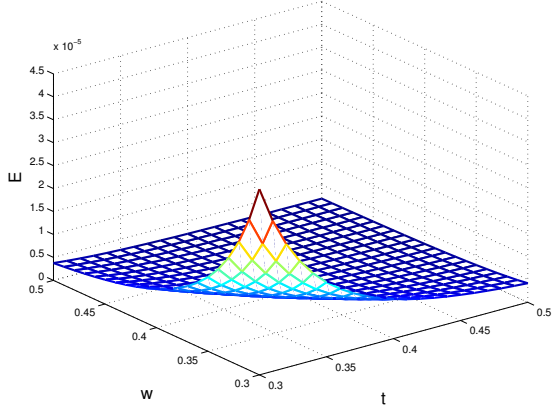}
       % \label{fig:CSG}
    }
    \subfigure[W = 0.5, t = 0.5 to W = 1, t = 1]{
        \includegraphics[scale=0.4]{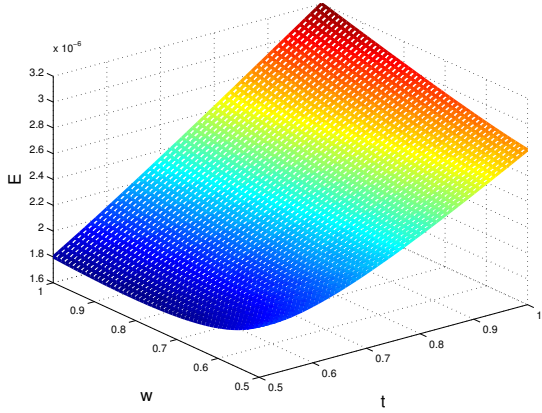}
       % \label{fig:OpenAccess}
    }

    \caption{Zooming-in pictures for Fig.~2. from W = 0.3, t = 0.3 to W = 0.5, t = 0.5 and from W = 0.5, t = 0.5 to W = 1, t = 1, respectively.}
    %\label{fig:AccessMethods}
\end{figure}

\begin{table}
\caption{Data for $t$ = 1 and $W$ = 1 in Fig.~2, respectively}
\centering
\subtable[Data for $t$ = 1 in Fig.~2.]{
  \begin{tabular}{|c|c|}
  \hline
  Fixed Delay & $t$ = 1  \\ \hline \hline
  $W$ = 0.1  & 2.2560e-05 \\ \hline
  $W$ = 0.2  & 3.4400e-06 \\ \hline
  $W$ = 0.3  & 2.8448e-06 \\ \hline
  $W$ = 0.4  & 2.7725e-06 \\ \hline
  $W$ = 0.5  & 2.8000e-06 \\ \hline
  $W$ = 0.6  & 2.8610e-06 \\ \hline
  $W$ = 0.7  & 2.9369e-06 \\ \hline
  $W$ = 0.8  & 3.0205e-06 \\ \hline
  $W$ = 0.9  & 3.0205e-06 \\ \hline
  $W$ = 1.0  & 3.2000e-06 \\ \hline
\end{tabular}
       \label{tab:firsttable}
}
\qquad
\subtable[Data for $w$ = 1 in Fig.~2.]{
  \begin{tabular}{|c|c|}
  \hline
  Fixed Bandwidth & $W$ = 1  \\ \hline \hline
  $t$ = 0.1  & 2.0760e-05 \\ \hline
  $t$ = 0.2  & 1.8400e-06 \\ \hline
  $t$ = 0.3  & 1.4448e-06 \\ \hline
  $t$ = 0.4  & 1.5725e-06 \\ \hline
  $t$ = 0.5  & 1.8000e-06 \\ \hline
  $t$ = 0.6  & 2.0610e-06 \\ \hline
  $t$ = 0.7  & 2.3369e-06 \\ \hline
  $t$ = 0.8  & 2.6205e-06 \\ \hline
  $t$ = 0.9  & 2.9088e-06 \\ \hline
  $t$ = 1.0  & 3.2000e-06 \\ \hline
\end{tabular}	
       \label{tab:secondtable}
}
\end{table}

For fixed $t$, the relationship between $E$ and $W$ is expressed as (setting $t$ = 1 for simplicity)
\begin{equation}
\label{E_WW}
\begin{aligned}
E = \frac{\left(2^{\frac{1}{W}}-1\right)WN_{0}}{g} + WP_{cir} + P_{sb}
\end{aligned}
\end{equation}

For fixed $W$, the relationship between $E$ and $t$ is expressed as (setting $W$ = 1 for simplicity)
\begin{equation}
\label{E_tt}
\begin{aligned}
E = \frac{\left(2^{\frac{1}{t}}-1\right)tN_{0}}{g} + tP_{cir} + tP_{sb}
\end{aligned}
\end{equation}

In Fig.~4 and Fig.~5, the three curves marked with TP-d=600m, TP-d=800m, and TP-d=1000m show the relationship between $E_{tran}$ and $W$ and $t$, respectively. The three curves marked with OP-d=600m, OP-d=800m, and OP-d=1000m show the relationship between $E$ and $W$ and $t$, respectively. Fig.~4 and Fig.~5 show that taking operating power into account, for fixed transmission distance $d$, there exist the minimum energies per bit for the given limited bandwidths and delays, respectively.
\begin{center}
\begin{figure}[!htb]
\includegraphics[width=3.3 in]{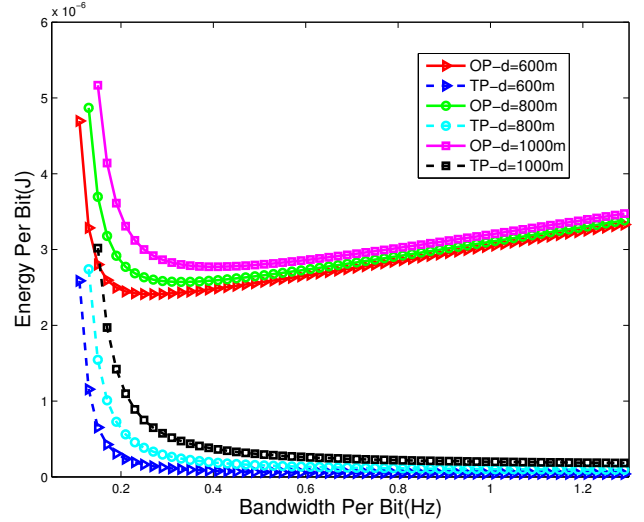}
\caption{Relation between energy per bit and bandwidth per bit.}
\end{figure}
\end{center}
\begin{center}
\begin{figure}[!htb]
\includegraphics[width=3.3 in]{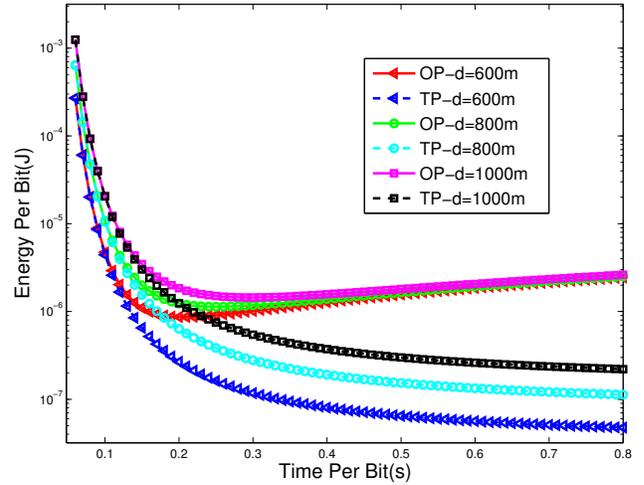}
\caption{Relation between energy per bit and delay time.}
\end{figure}
\end{center}

\subsection{Optimal Tradeoff Strategies between Energy and Bandwidth/Delay}
Taking the derivative over Eq.~\eqref{E_WW} with respect to $W$ and setting the result to zero, we can obtain
\begin{equation}
\label{E_W_Deri}
\begin{aligned}
-\frac{N_{0}2^{\frac{1}{W}}\log2}{Wg} + \frac{\left(2^{\frac{1}{W}}-1\right)N_{0}}{g} + P_{cir} = 0
\end{aligned}
\end{equation}
Simplifying Eq.~\eqref{E_W_Deri}, we can obtain
\begin{equation}
\label{W-optimal}
\begin{aligned}
2^{\frac{1}{W_{o}}}\left(1-\frac{\log2}{W_{o}}\right) = \frac{N_{0}-gP_{cir}}{N_{0}}
\end{aligned}
\end{equation}
where $W_{o}$ is the optimal bandwidth for the minimum energy per bit.

%Using Taylor's formula for $f(x) = 2^{x}$, equation (10) can be reduced as
%\begin{equation}
%\begin{aligned}
%x^{2}\mathrm{ln}2 + x\mathrm{ln}2 +A-2 = 0
%\end{aligned}
%\end{equation}
%
%The result for equation (11) is
%\begin{equation}
%\begin{aligned}
%x = \cfrac{-\mathrm{ln}2 + \sqrt{\mathrm{ln}^{2}2-4(A-2)\mathrm{ln}2}}{2\mathrm{ln}2}
%\end{aligned}
%\end{equation}
%
%So the optimal bandwidth $W_\mathrm{optimal}$ is
%\begin{equation}
%\begin{aligned}
%W_\mathrm{optimal} = \cfrac{2\mathrm{ln}2}{-\mathrm{ln}2 + \sqrt{\mathrm{ln}^{2}2-4(A-2)\mathrm{ln}2}}
%\end{aligned}
%\end{equation}

Taking the derivative over Eq.~\eqref{E_tt} with respect to $t$, we can obtain
\begin{equation}
\label{E_t_Deri}
\begin{aligned}
-\frac{N_{0}2^{\frac{1}{t}}\log2}{tg} + \frac{\left(2^{\frac{1}{t}}-1\right)N_{0}}{g} + P_{cir} + P_{sb} = 0
\end{aligned}
\end{equation}
Simplifying Eq.~\eqref{E_t_Deri}, we can obtain
\begin{equation}
\label{t-optimal}
\begin{aligned}
2^{\frac{1}{t_{o}}}\left(1-\frac{\log2}{t_{o}}\right) = \frac{N_{0}-g(P_{cir}+P_{sb})}{N_{0}}
\end{aligned}
\end{equation}
where $t_{o}$ is the optimal $t$ for the minimum energy per bit.

%Using $B$ and $x$ represent $\cfrac{N_{0}-g(P_{cir}+P_{cons})}{N_{0}}$ and $\cfrac{1}{t}$ respectively. And also using Taylor's formula for $f(x) = 2^{x}$. Like mathmatically manipulations for $W_\mathrm{optimal}$, we can obtain the optimal delay time
%\begin{equation}
%\begin{aligned}
%t_\mathrm{optimal} = \cfrac{2\mathrm{ln}2}{-\mathrm{ln}2 + \sqrt{\mathrm{ln}^{2}2-4(B-2)\mathrm{ln}2}}
%\end{aligned}
%\end{equation}

Under the free space propagation model, for fixed $G_{t}$, $G_{r}$, $\lambda$, and $L$, the channel gain, denoted by $g(d)$, is the function of transmission distance $d$ which can be expressed as
\begin{equation}
\begin{aligned}
g(d)=\frac{G_{t}G_{r}\lambda^2}{(4\pi)^2 d^\alpha L}
\end{aligned}
\end{equation}
where $G_{t}$ and $G_{r}$ are the transmit and receive antenna gains, respectively, $\lambda$ is the wavelength, and $L$ is the system loss unrelated to propagation ($L\ge $1).

Therefore, for fixed $d$, there exists the optimal $W$ and $t$ for the minimum energy per bit, which are $W_\mathrm{o}$ and $t_\mathrm{o}$. Fig.~4 and Fig.~5 show that the curves for Eq.~\eqref{E_WW} and Eq.~\eqref{E_tt} are convex. Thus there are the unique solutions for $W_\mathrm{o}$ and $t_\mathrm{o}$, respectively. The optimal strategy for the minimum energy per bit is to choose different $W_\mathrm{o}$ and $t_\mathrm{o}$ pairs for different transmission distances adaptively. However, in real wireless networks, the available bandwidth can be smaller or larger than $W_\mathrm{o}$, and the acceptable delay can be also smaller or larger than $t_\mathrm{o}$. Then, our next problem is how to use the available bandwidth or the acceptable delay trading for saving energy? For this purpose, we develop Algorithms 1 and 2 to search for optimal $W_\mathrm{o}$ and $t_\mathrm{o}$, respectively. Methods in Algorithms 1 and 2 are similar, but for searching $W_\mathrm{o}$ and $t_\mathrm{o}$, respectively.

\renewcommand{\algorithmicrequire}{\textbf{Input:}}
\renewcommand{\algorithmicensure}{\textbf{Output:}}

\begin{algorithm}[h]         %??・¨μ??aê?
\caption{ Adaptive $W_\mathrm{o}$ Strategy.}             %??・¨μ?±êìa
\label{alg:Framwork}                  %????・¨ò???±ê??￡??a?ù・?±??ú???D????・¨μ?òyó?
\begin{algorithmic}[1]                %2??a[1]ê??é??μ?￡?
\REQUIRE ~~$W_\mathrm{a}$                         %??・¨μ?ê?è?2?êy￡oInput

\ENSURE ~~$W_\mathrm{user}$                           %??・¨μ?ê?3?￡oOutput

  \FOR {i = 1:UN}
    \STATE  Calculate $d_\mathrm{i}$
    \STATE  Calculate $W_\mathrm{o}(d_\mathrm{i})$ by Eq.~\eqref{W-optimal}
  \IF {$W_\mathrm{a} > W_\mathrm{o}(d_\mathrm{i})$}
    \STATE $W_\mathrm{user}(d_\mathrm{i})$ = $W_\mathrm{o}(d_\mathrm{i})$
  \ELSE
    \STATE $W_\mathrm{user}(d_\mathrm{i})$ = $W_\mathrm{a}$
  \ENDIF
  \ENDFOR

%\label{code:fram:extract}
%\RETURN $E_n$;                %??・¨μ?・μ???μ
\end{algorithmic}
\end{algorithm}

\begin{algorithm}[h]         %??・¨μ??aê?
\caption{ Adaptive $t_\mathrm{o}$ Strategy.}             %??・¨μ?±êìa
\label{alg:Framwork}                  %????・¨ò???±ê??￡??a?ù・?±??ú???D????・¨μ?òyó?
\begin{algorithmic}[1]                %2??a[1]ê??é??μ?￡?
\REQUIRE ~~$t_\mathrm{a}$                         %??・¨μ?ê?è?2?êy￡oInput

\ENSURE ~~$t_\mathrm{user}$                           %??・¨μ?ê?3?￡oOutput

  \FOR {i = 1:UN}
    \STATE  Calculate $d_\mathrm{i}$
    \STATE  Calculate $t_\mathrm{o}(d_\mathrm{i})$ by Eq.~\eqref{t-optimal}
  \IF {$t_\mathrm{a} > t_\mathrm{o}(d_\mathrm{i})$}
    \STATE $t_\mathrm{user}(d_\mathrm{i})$ = $t_\mathrm{o}(d_\mathrm{i})$
  \ELSE
    \STATE $t_\mathrm{user}(d_\mathrm{i})$ = $t_\mathrm{a}$
  \ENDIF
  \ENDFOR

%\label{code:fram:extract}
%\RETURN $E_n$;                %??・¨μ?・μ???μ
\end{algorithmic}
\end{algorithm}

In Algorithm 1 and 2, UN and $d_{i}$ denote the user number in the wireless networks and the distance between the BS and user $i$. $W_\mathrm{a}$ and $t_\mathrm{a}$ represent the available bandwidth and acceptable delay for each bit, respectively. $W_\mathrm{user}(d_{i})$ and $t_\mathrm{user}(d_{i})$ represent the bandwidth and the delay which the user $i$ should trade for the minimum energy per bit. $W_\mathrm{o}(d_{i})$ and $t_\mathrm{o}(d_{i})$ represent the optimal bandwidth and delay for user $i$.

%However, it is not easy to obtain the analytical value of $W_\mathrm{o}$ and $t_\mathrm{o}$ from equation (10) and (12). We use step $W_\mathrm{o}$ and $t_\mathrm{o}$ values for different $d$. In each region, we adjust the $W$ to $W_\mathrm{o}$ or adjust $t$ to $t_\mathrm{o}$. In Table 1, we give out the step values of $W_\mathrm{o}$ and $t_\mathrm{o}$ by using numerical method.
%\begin{table}
%\begin{center}
%\caption{Optimal Delay Time and Bandwidth in Different Range}
%\begin{tabular}{c|c|c}
%  \hline
%  d(m) & $t_\mathrm{o} $(s)&$W_\mathrm{o} $(Hz)\\ \hline
%  0-100    & 0.09 & 0.102 \\ \hline
%  100-200  & 0.12 & 0.138 \\ \hline
%  200-300  & 0.14 & 0.170 \\ \hline
%  300-400  & 0.16 & 0.201 \\ \hline
%  400-500  & 0.18 & 0.233 \\ \hline
%  500-600  & 0.21 & 0.265 \\ \hline
%  600-700  & 0.23 & 0.298 \\ \hline
%  700-800  & 0.25 & 0.332 \\ \hline
%  800-900  & 0.27 & 0.367 \\ \hline
%  900-1000 & 0.30 & 0.404 \\ \hline
%\end{tabular}	
%\end{center}
%\end{table}

%Adding two dimension optimization here.

\section{Simulation Results}
\label{Simulation Results}
To evaluate our adaptive strategies, we consider a classical hexagonal deployment which is shown in Fig.~6.
\begin{center}
\label{Hexagon}
\begin{figure}[!htb]
\includegraphics[width=3.3 in]{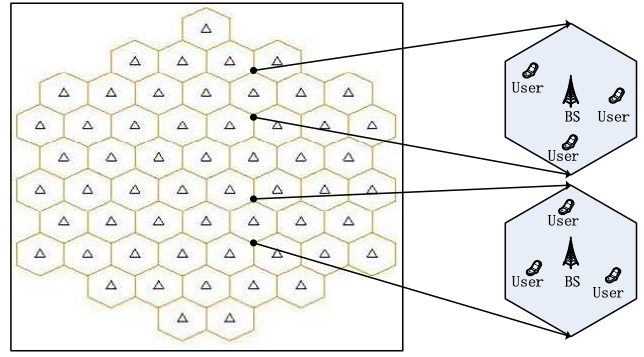}
\caption{Hexagonal network deployment.}
\end{figure}
\end{center}

As shown in Fig.~6, the given 57 cells are deployed and the radius of each cell is 1000m. For each cell $C_{i}$ (1$\le$ i $\le$ 57), $B_{i}$ and $U_{i}$ denote the base station (BS) and mobile user (MU) in each cell, respectively. $d_{i}$ denotes the current distance from $B_{i}$ to $U_{i}$. The data sent by base station $B_{i}$ is transmitted with power $P_{i}$. The system bandwidth is 20MHz and the total number of users is 500. Because different wireless networks have different available bandwidths and acceptable delays, we will show the relationship between energy consumption and available bandwidth, the relationship between energy consumption and acceptable delay, respectively. The related simulation parameters are listed in TABLE II.
\begin{table}
\begin{center}
\caption{Simulation Parameters}
\begin{tabular}{c|c}
  \hline
  Parameter & Value \\\hline  \hline
  Carrier Frequency $f_c$ & 2.4GHz \\\hline
  Cell Radius $R$ & 1000m \\\hline
  Transmit Antenna Gain $G_t$ & 1 \\\hline
  Receiver Antenna Gain $G_r$ & 1 \\\hline
  Circuit Power $P_{cir}$ & $1\times 10^{-6}W/Hz$\\\hline
  Static Power $P_{sb}$ & $2\times 10^{-6}W$\\\hline
  System Loss L &   2.5 \\\hline
  PSD of The Local Noise $N_0$ & $8 \times 10^{-21}$\\\hline
  Path-Loss Exponent  $\alpha$ & 3\\\hline
\end{tabular}
\end{center}
\end{table}

\begin{center}
\begin{figure}[!htb]
\includegraphics[width=3.3 in]{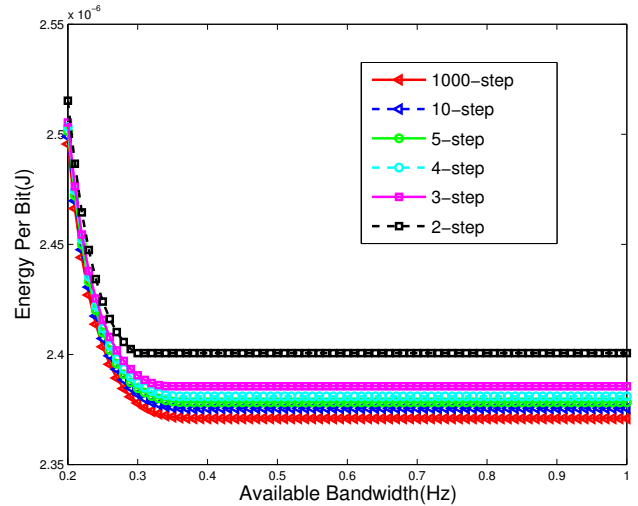}
\caption{Energy per bit versus available bandwidth (n-step in the legend indicates that n can be 2, 3, 4, 5, 10, and 1000).}
\end{figure}
\end{center}
\begin{center}
\begin{figure}[!htb]
\includegraphics[width=3.3 in]{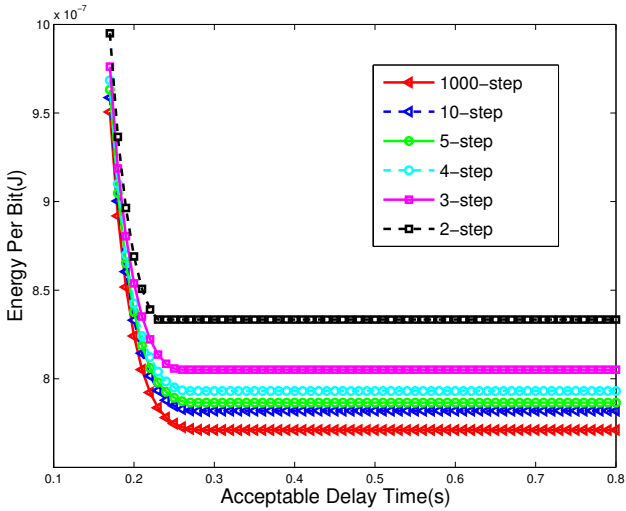}
\caption{Energy per bit versus acceptable delay time (n-step in the legend indicates that n can be 2, 3, 4, 5, 10, and 1000).}
\end{figure}
\end{center}
Fig.~7 shows that the energy per bit versus available bandwidth with optimal bandwidths and different steps. The optimal adaptive strategy uses the continuous value for $W_\mathrm{o}$. This implies that for different $d$, the system uses different optimal bandwidths $W_\mathrm{o}$ obtained by Algorithm 1. In Fig.~7, n-step represents that n types of bandwidths, denoted by $W_\mathrm{o}(i)$ (1 $\le$ $i$ $\le$ n), can be used in the wireless networks. $W_\mathrm{o}(i)$ (1 $\le$ $i$ $\le$ n) can be calculated using Algorithm 1. When n is fixed, the users located in the $i$-th step use the same bandwidth $W_\mathrm{o}(i)$ (1 $\le$ $i$ $\le$ n). For instance, n = 5, the used bandwidth $W_\mathrm{user}$ for the user located in the 1000($i$-1)/n - 1000$i$/n m is $W_\mathrm{o}(i)$ (1 $\le$ $i$ $\le$ 5). Fig.~7 shows that when available bandwidth increases, the energy per bit decreases in the case of small available bandwidth. However, the energy per bit finally decreases to one constant, which is the minimum energy per bit for the used strategy. This is because when the available bandwidth is small, most users cannot use the optimal bandwidth for them. When the available bandwidth becomes larger, more users can use their optimal bandwidths. When all users' optimal bandwidths are smaller than the available bandwidth, the energy per bit decreases to the constant. The larger n, the smaller energy per bit. This is because the larger n, the smaller difference between $W_\mathrm{o}(i)$ (1 $\le$ $i$ $\le$ n) and $W_\mathrm{o}$. Hence more users can make use of their optimal bandwidths, decreasing the energy per bit. Fig.~7 also shows that the energy per bit does not decrease when available bandwidth is large to some extent. For example, with 2-step, when available bandwidth is larger than 0.3Hz, the energy per bit remains as 2.4e-6J.

Fig.~8 shows the curves of the energy per bit versus acceptable delay with optimal delays and different steps. The optimal adaptive strategy uses the continuous value for $t_\mathrm{o}$. This implies that for different $d$, the system uses the optimal delays $t_\mathrm{o}$ obtained by Algorithm 2. In Fig.~8, n-step represents that n types of delays, denoted by $t_\mathrm{o}(i)$ (1 $\le$ $i$ $\le$ n), can be used in the wireless networks. $t_\mathrm{o}(i)$ (1 $\le$ $i$ $\le$ n) can be calculated using Algorithm 2. When n is fixed, the users located in the $i$-th step use the same delay $t_\mathrm{o}(i)$ (1 $\le$ $i$ $\le$ n). For instance, n = 5, the used delay $t_\mathrm{user}$ for the user located in the 1000($i$-1)/n - 1000$i$/n m is $t_\mathrm{o}(i)$ (1 $\le$ $i$ $\le$ 5). Fig.~8 shows that when acceptable delay increases, the energy per bit decreases to one constant, which is the minimum energy per bit. This is because when the acceptable delay is small, most users cannot wait for the optimal delay. Then the acceptable delay becomes larger, more users can use their optimal delay. When all users' optimal delay are smaller than the acceptable delay time, the energy per bit decreases to the constant. The larger n, the smaller energy per bit. This is because the larger n, the smaller difference between $t_\mathrm{o}(i)$ (1 $\le$ $i$ $\le$ n) and $t_\mathrm{o}$. Hence more users can make use of their optimal delay, decreasing the energy per bit. Fig.~8, also shows that the energy per bit does not decrease when acceptable delay is large to some extent. For example, with 2-step, when acceptable delay is larger than 0.22s, the energy per bit remains as 8.3e-7J.

The minimum energies per bit in Fig.~7 and in Fig.~8 are different. This is because that we employ the fixed bandwidth for optimal delay and the fixed delay time for optimal bandwidth, respectively.

\section{Conclusions}
\label{Conclusions}
We proposed the resource trading strategy for On-Demand performance in wireless networks. Under demanded performance constraint, any wireless resource can be saved by consuming the other wireless resources instead. For green wireless networks, we proposed the optimal resource trading for bandwidth-energy trading and time-energy trading regardless of RAL. Based on the best resource trading, we developed the adaptive strategy by using related $W_\mathrm{user}(d)$ or $t_\mathrm{user}(d)$ for different transmission distances in wireless networks. The larger number of steps used to obtain $W_\mathrm{o}$ or $t_\mathrm{o}$, the less energy resource will be consumed.

\end{document}